\documentclass[aps,twocolumn,showpacs,preprintnumbers]{revtex4-1}

\usepackage{graphicx}  
\usepackage{subfigure}
\usepackage{multirow}
\usepackage[final]{changes}
\usepackage{ulem}
\usepackage{color}
\usepackage{fancyhdr}
\usepackage{longtable}
\usepackage{parskip}
\usepackage[T1]{fontenc}
\usepackage{dcolumn}   
\usepackage{physics}
\usepackage{bm}        
\usepackage{amsfonts}  
\usepackage{amsmath}   
\usepackage{amssymb}   
\usepackage{url}
\usepackage{hyperref}
\hypersetup{
    colorlinks=true,
    linkcolor=blue,
    filecolor=blue,
    urlcolor=blue,
    citecolor=blue,
}

\newcommand{\pwisein}{\left\{ \begin{array}{ll}}
\newcommand{\pwiseout}{\end{array}\right.}

\newcommand{\bi}{\boldsymbol}

\newcommand{\ii}{\mathrm{i}}
\newcommand{\ee}{\mathrm{e}}
\begin{document}

\title{Quantum Zeno Effect in Heisenberg Picture and Critical Measurement Time}

\author{Wu Wang\textsuperscript{1}}

\author{Ruo-Xun Zhai\textsuperscript{2}}

\author{C. P. Sun\textsuperscript{1,2,}}

\email[\textsuperscript{}]{cpsun@csrc.ac.cn}

\affiliation{\textsuperscript{1}Beijing Computational Science Research Center, Beijing 100193, China \\
\textsuperscript{2}Graduate School, China Academy of Engineering Physics, Beijing 100193, China}

\date{March 22, 2021}

\begin{abstract}
Quantum Zeno effect is conventionally interpreted by the assumption of the wave-packet collapse, in which does not involve the duration of measurement. However, we predict duration $\tau_m$ of each measurement will appear in quantum Zeno effect by a dynamical approach. Moreover, there exists a model-free critical measurement time, which quantum Zeno effect does not occur when $\tau_m$ takes some special values. In order to give these predictions, we first present a description of quantum Zeno effect in the Heisenberg picture, which is  based on the expectation value of an observable and its fluctuation. Then we present a general proof for quantum Zeno effect in the Heisenberg picture, which is independent of the concrete systems. Finally, we calculate the average population and relative fluctuation after $N$ successive measurements in XX model, which agrees with our prediction about the critical measurement time.
\end{abstract}


\maketitle

\section{Introduction}
 Any unstable state of a quantum system will jump to other states with the time evolution, but this transition can be inhibited by performing frequent measurements. Known as the quantum Zeno effect (QZE) \cite{Misra_1977_JMP}, this
phenomenon has been demonstrated to support the  von Neumann's postulate\textemdash wave-packet collapse (WPC) \cite{von_mathqm}. However, many authors  pointed out that the WPC is not the unique way to  explain the QZE. For example,
Peres suggested that a modified Hamiltonian slows down the decay of an unstable state
in a two-level system \cite{Peres_1980}. It is noted that the argument for this resolution was carried out in the Schr\"{o}dinger picture without using the WPC. When the QZE in atomic transition was reported by
 Itano \textit{et al.} \cite{Itano_1990_PRA} and interpreted as a witness of the WPC,  Petrosky \textit{et al.} understood the QZE as a dynamical process without appealing to  the WPC. Their proposals could also recover Itano \textit{et al.}'s experiment \cite{Petrosky_1990_PLA,Petrosky_1991_PA}. Later on, Sun \textit{et al.} suggested a general dynamical
 approach to the QZE \cite{Sun_1995_QZE_FDP}. Ai \textit{et al.} studied the QZE  phenomenon via the dynamical method based on "quasi-measurement"\textemdash which is substantially pre-measurement and leads to a entanglement between system and  apparatus. \cite{Ai_2013}. Harrington \textit{et al.} subsequently realized quasi-measurement in a superconducting qubit experiment and observed the suppression of the qubit decay \cite{Harrington_PRL_2017}.

In 2008, Bernu \textit{et al.} observed suppression of photon number in a cavity quantum electrodynamics (CQED)
experiment and interpreted it with the WPC \cite{Bernu_PRL_2008}. Xu \textit{et al.}, however, demonstrated their results could be also interpreted via the unitary evolution of a dynamical model \cite{Xu_PRA_2011}. {Later, Raimond \textit{et al.} also interpreted the CQED experiment by dynamical method, and further proposed  to realize QZE-like phenomenon\textemdash quantum Zeno dynamics via a CQED experiment \cite{Raimond_PRL_2010}. It is found by ref.\cite{Xu_PRA_2011} that when quantum measurement process is interpreted in dynamical unitray evolution, the interaction time or measurement time \(\tau_m\) will appear in the result, which never appears in the interpretation of the WPC. Besides, they find that there exists a critical measurement time $\tau_m$, where the QZE will not occur. Later,  Zheng \textit{et al.}'s experiment verified the critical measurement time in nuclear magnetic resonance systems \cite{Zheng_PRA_2013_QZE}.

Now, whether the critical measurement time  universally exists in the practical QZE still remains to be proved. This paper is ready to answer this question.
To this end, we first investigate how to describe QZE only through the
measurement results. From the perspective of physical observation, what we can measure
is only the expectation value and fluctuation correlation of dynamical variables,  which can be calculated conveniently in the Heisenberg picture. Therefore, we generally present the universal description of QZE in the Heisenberg picture. What we would like to show is that
frequent measurements turn the expectation value of an observable into a constant and its fluctuation to zero. With this observation, we obtain a universal proof for QZE in the Heisenberg picture, especially a  model-free  critical measurement time.
Without loss of generality, measurement is dynamically described by the nondemolition interaction between the system and the apparatus.

This article is organized as follows: In Sec. \ref{sec_2020_10_22_1}, we first recall QZE phenomenon
in Schr\"{o}dinger picture, then reformulate it via the expectation value  of an observable and its fluctuation. In Sec. \ref{sec_2020_10_22_2},
we describe the dynamical process of QZE in the Heisenberg picture with respect to dynamical variables. In Sec. \ref{sec_2020_10_22_3}, we explore that under what conditions
 the QZE phenomenon occurs, and under these conditions, we introduce the concept of universal critical measurement time.
In Sec. \ref{sec_2020_10_22_4}, we calculate the average
population and relative fluctuation after $N$ successive measurements in XX model, and discuss the numerical results. Finally, a conclusion is
given in Sec. \ref{sec_2020_10_22_5}.

 For convenience, we set $\hbar=1$ in the following discussion.

\section{\label{sec_2020_10_22_1}Revisiting QZE via measurement of expectation value  }

Let us first recall the dynamical description of the QZE in the Schr\"{o}dinger picture \cite{Xu_PRA_2011}.
Suppose we have a system $S$ and an apparatus $E$, the free evolution of the system can be
described by a unitary evolution operator $U(t)=\mathrm{exp}({-iHt})$, where $H=H_0+V$ is the Hamiltonian of $S$. $H_0$ and $V$ are the free Hamiltonian and the interaction that causes transitions of system states, respectively. A measurement can be given by another unitary evolution operator $M(t)=\mathrm{exp}(-iH_{\mathrm{M}}t)$, where $H_\mathrm{M}$ is the couplings of system $S$ to the apparatus $E$. Here, we consider the interaction Hamiltonian $H_M$ dominates the Hamiltonian of the total system $S$ plus $E$. Usually, we choose
$[H_0, H_M]=0$ which makes $M(t)$ to satisfy a quantum  nondemolition (QND) measurement \cite{Braginsky_1996_RMP}. Thus, by assuming $\ket{s_i}$ is an eigenstate of the free Hamiltonian $H_0$ and $\ket{e}$ is initial state of apparatus $E$, a measurement is described by a mapping
\begin{equation}\label{2020_10_18_1}
 M(t): ~ \ket{s_i}\otimes\ket{e}\rightarrow  \ket{s_i}\otimes\ket{e_i}~,
\end{equation}
where $\ket{e_i}$ is a state of $E$ corresponds to the state $\ket{s_i}$ of $S$. For an ideal measurement, $\ket{e_i}$ is orthonormal with each other, i.e., $
\langle e_i|e_{j}\rangle=\delta_{ij}$, which leads to the measurement results could be well distinguished. Now, the QZE phenomenon with $N$ frequent measurements {(see Fig.\ref{Fig:model_2020_12_18_1})} is described via a unitary evolution operator
\begin{equation}\label{2020_10_18_2}
  U(\tau,\tau_m)=\big[M(\tau_m)U(\tau)\bi]^N~,
\end{equation}
where $\tau$ is a small duration of each free evolution, and $\tau_m$ a fixed time interval of each  measurement. The relationship between $\tau$ and $N$ is $t=N\tau$, which
means the total duration of the free evolution is fixed as $t$. In the large-$N$ limit,  $N\rightarrow\infty$, we claim that if $U(\tau,\tau_m)$ becomes diagonal with respect
 to basis $\ket{s_i}$, then the QZE phenomenon occurs. $U(\tau,\tau_m)$ becomes diagonal
 means if the initial state of $S$ is $\ket{s_i}$, the final state will remain $\ket{s_i}$,
thus this is in agreement with conventional description of the QZE with an interpretation in terms of the WPC. In the large-$N$ limit, this phenomenon will happen indeed \cite{Xu_PRA_2011}.
\begin{figure}[t]
\includegraphics[width=3.5in]{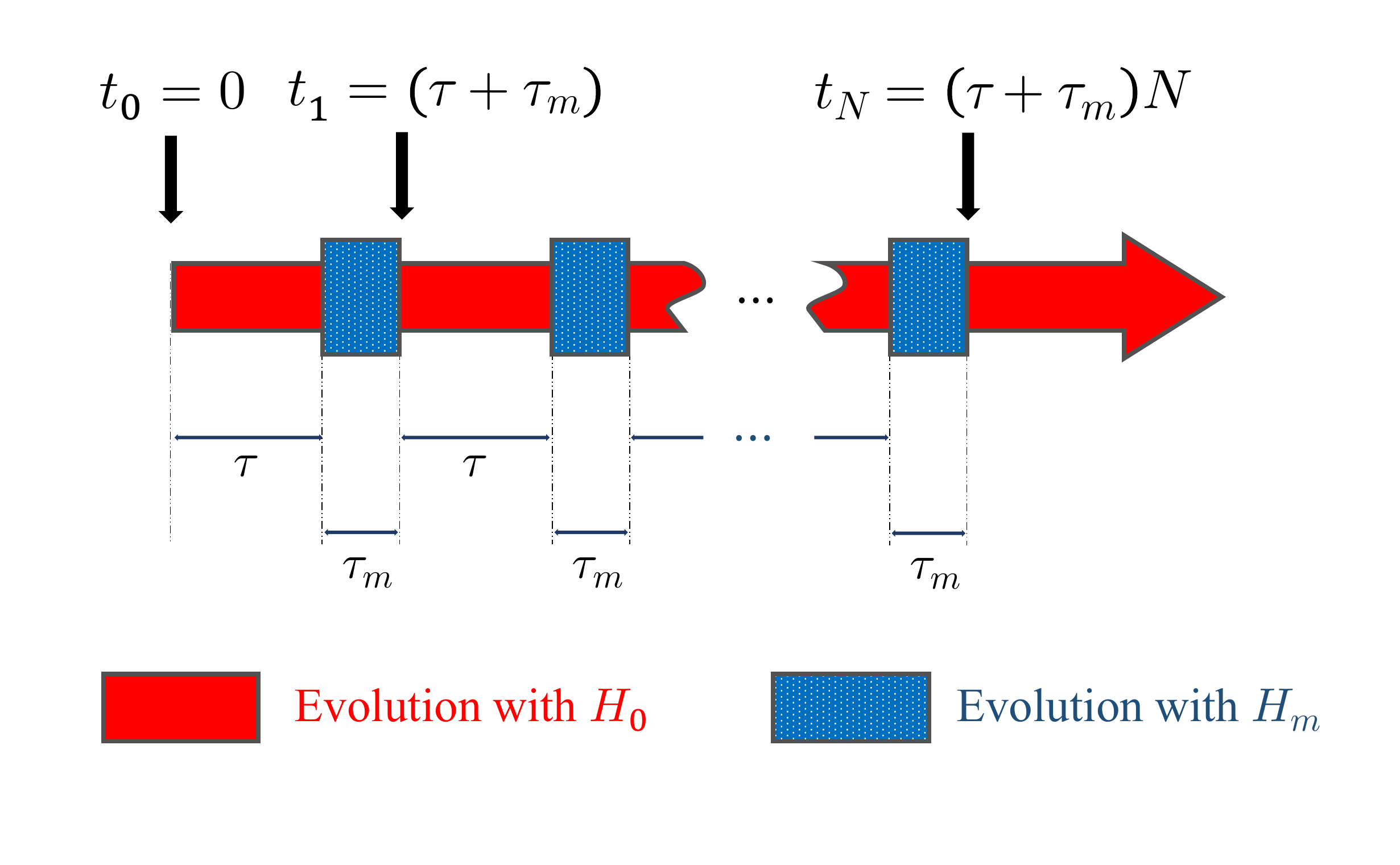}
\caption{\label{Fig:model_2020_12_18_1}  { Dynamical evolution process with $N$ successive measurements.  Each measurement (blue block) is followed by a free evolution (red block). Time interval of each measurement is $\tau_m$, and duration of each free evolution $\tau$}.
}
\end{figure}
However, what we can actually  directly observe from the experiment is the expectation value
of a system's operator ${A}$ rather than  the state of the system. Thus we need to revisit
the QZE phenomenon from a perspective with respect to the expectation value.
Therefore, let us firstly see a simple proposition about the expectation value:  the
fluctuation of an operator ${A}$ of a system is zero, if and only if the state of
the system is the eigenstate of ${A}$ with egienvalue $\bar{A}$. Here, $\bar{A}$
is the  expectation value of ${A}$.  This is known from the identity
\begin{equation}\label{2020_10_18_3}
  \Delta^2A=\langle ({A}-\bar{A})\psi| ({A}-\bar{A})\psi\rangle~,
\end{equation}
where $\ket{\psi}$ is the state of system. Supposing the initial states of system $S$ and
  apparatus $E$ are $\rho_S(0)$ and $\rho_E(0)$, respectively, we write the expectation
   value of ${A}$ at time $t_N=(\tau+\tau_m)N$ as
\begin{equation}\label{2020_10_19_1}
  \bar{A}(t_N)=\mathrm{Tr}\big({A}U(\tau,\tau_m)\rho_S(0)\otimes\rho_E(0)U(\tau,\tau_m)^\dagger\big)~.
\end{equation}
Thus, by this proposition we claim that if
\begin{equation}\label{2020_10_19_2}
   \bar{A}(t_N)\rightarrow \bar{A}(0),~~~ \Delta^2A(t_N)\rightarrow0
\end{equation}
in the large-$N$ limit, then the state of the system will be $\rho_S(0)$. Here,
$\bar{A}(0)$ is the initial expectation value of ${A}$ given
 by $\mathrm{Tr}\big({A}\rho_S(0)\otimes\rho_E(0)\big)$, {and we assume $H_M$ is  not degenerate. } With this observation,
  we use Eq.\eqref{2020_10_19_2} instead of disappearance of $U(\tau,\tau_m)$'s
  off-diagonal parts in the large-$N$ limit to describe the QZE phenomenon.

\section{\label{sec_2020_10_22_2}QZE in the Heisenberg picture }

  We have reformulated QZE phenomenon via the expectation value: The expectation value of an observable and its fluctuation turn into a constant and zero respectively, due to frequent measurements. It is convenient to calculate the expectation value of an observable in the Heisenberg picture. Thus we now choose the Heisenberg picture to investigate QZE phenomenon.

Suppose the initial Hamiltonian of the system $S$ and the initial interaction Hamiltonian between $S$ and  apparatus $E$ are $H=H_0+V$ and $H_M$ respectively. Here, $V$ causes transitions among the states of the system $S$, which satisfies $[H_0,V]\neq0$. We choose $[H_0,{H_M}]=0$ to satisfy the QND measurement condition \cite{Braginsky_1996_RMP}. If we wish to measure a dynamical variable ${A}$, $A$ should satisfy conditions $[H_0,A]=0$ and $[{H_M},A]=0$. Then, we describe the dynamical process of $N$ successive measurements (see Fig.\ref{Fig:model_2020_12_18_1}) for variable $A$ in the Heisenberg picture as
\begin{equation}\label{2020_10_20_1}
  i{\dot{A}}(t)=[{A}(t),V(t)]
\end{equation}
when $t_k\leq t<t_k+\tau$, and
\begin{equation}\label{2020_10_20_2}
  i{\dot{A}}(t)=0
\end{equation}
when $t_k+\tau\leq t<t_{k+1}$. Eq.\eqref{2020_10_20_2} means $A$ does not evolute with $H_M$ when measurements are carried out. We have denoted $t_k$ by $(\tau+\tau_m)k$, and $k$ is an
integer that satisfies $0<k\leq N$. The
total duration of free evolution is still fixed as $t=N\tau$. In order to form closed
 equations, we write down the Heisenberg equations of motion about other variables as follows:
\begin{equation}\label{2020_10_20_3}
\left\{\begin{split}
   i{\dot{V}}(t) &  =[{V}(t),H(t)]=[{V}(t),H_0(t)]~, \\
    i{\dot{H}_0}(t) &  =[{H_0}(t),H(t)]=[H_0(t),{V}(t)]~, \\
    i{\dot{H}_M}(t) & =[{H_M}(t),H(t)] =[{H_M}(t),V(t)]~,
\end{split}\right.
\end{equation}
when $t_k\leq t<t_k+\tau$, and
\begin{equation}\label{2020_10_20_4}
\left\{\begin{split}
   i{\dot{V}}(t) &  =[V(t),H_M(t)] ~,\\
   i\dot{H}_0(t) &  =0 ~,\\
    i{\dot{H}_M}(t) & =0~,
\end{split}\right.
\end{equation}
when $t_k+\tau\leq t<t_{k+1}$. In the above arguments, there is an implied assumption
that the interaction Hamiltonian dominates the total Hamiltonian when
measurements are carried out.

 Next, {we can claim
 that if the initial state of the system is an eigenstate of $H_0$}, and the solution $A(t_N)$ corresponding to Eq.\eqref{2020_10_20_1} and
  Eq.\eqref{2020_10_20_2} satisfies
 \begin{equation}\label{2020_10_20_5}
   A(t_N)\rightarrow A(t_0)
 \end{equation}
in the large-$N$ limit, then the QZE  phenomenon occurs. This is obvious if one
notes that observable $A^2(t_N)$ will also tend to $A^2(t_0)$. Thus, the expectation value and  fluctuation will turn into a constant and zero, respectively.  In the next section, we will show that Eq.\eqref{2020_10_20_5} indeed is satisfied due to frequent measurement. From Eq.\eqref{2020_10_20_5}, we see that
$[V,{H_M}]\neq0$ must be satisfied, otherwise, the frequent measurements can not affect
the evolution of operator $A$ and no QZE phenomenon can occur. Thus,  we assume
$[V,{H_M}]\neq0$ in the following discussion.

\section{\label{sec_2020_10_22_3}Universal critical measurement time}
We next answer the question what can make the Eq.\eqref{2020_10_20_5} satisfy.
 The recursion relation of $A(t_k)$
 reads as
  \begin{equation}\label{2020_10_20_6}
    A(t_k)=-i\int_{t_{k-1}}^{t_{k-1}+\tau}[A(s),V(s)]\mathrm{d}s+A(t_{k-1})
  \end{equation}
  in accordance with the Eq.\eqref{2020_10_20_1} and Eq.\eqref{2020_10_20_2}.
  It follows from the recursion relation that
  \begin{equation}\label{2020_10_20_7}
    A(t_N)=-i\sum_{k=1}^{N}\int_{t_{k-1}}^{t_{k-1}+\tau}[A(s),V(s)]\mathrm{d}s+A(t_0)~.
  \end{equation}
Thus, for a short $\tau$ or a large $N$ with fixed $t$, $A(t_N)$ is approximated as
\begin{equation}\label{2020_10_20_8}
  A(t_N)\simeq-i\tau\sum_{k=1}^{N}[A,V](t_{k-1})+A(t_0)~.
\end{equation}
For convenience, we denote $[A,V](t)$ by $X(t)$, and its Heisenberg
equations of motion
is described from Eqs.\eqref{2020_10_20_3} and Eqs.\eqref{2020_10_20_4}   as
\begin{equation}\label{2020_10_22_1}
 \left\{ \begin{split}
    i\dot{X}(t) & =[X(t),H(t)] ~,~~t_k\leq t<t_k+\tau~,\\
     i\dot{X}(t)  &=[X(t),H_M(t)]~,~~t_k+\tau\leq t<t_{k+1}~.
  \end{split}\right.
\end{equation}
Similarly, the recursion relation of $X(t_k)$ via Eq.\eqref{2020_10_22_1} gives
\begin{equation}\label{2020_10_22_3}
   \begin{split}
    X(t_k) & =-i\sum_{l=0}^{k-1}\int_{t_{l}+\tau}^{t_{l+1}}[X(s),H_M(s)]\mathrm{d}s-\\
      & ~~~~i\sum_{l=0}^{k-1}\int_{t_{l}}^{t_{l}+\tau}[X(s),H(s)]\mathrm{d}s+X(t_{0})~.
  \end{split}
\end{equation}
Denoting $X(t_k)$ to the zero-order with respect to $\tau$ by $X(k\tau_m)$, we obtain
\begin{equation}\label{2020_10_22_4}
  X(k\tau_m)=-i\int_0^{k\tau_m}[X(s),{H_M(s)]}\mathrm{d}s+X(t_0)~.
\end{equation}
On the base vectors of $H_0$, ${H_M}$ and $X$ are expressed as
\begin{equation}\label{2020_10_22_5}
 \left\{ \begin{split}
   {H_M}(t_0) &= \sum_{n}{H(n)}|s_n\rangle\langle s_n|~, \\
     X(t_0) & =\sum_{mn}x_{nm}|s_n\rangle\langle s_m|~,
  \end{split}\right.
\end{equation}
where $\ket{s_n}$ is a common  eigenstate of $A$, $H_0$ and {$H_M$}. {$H(n)$ is a Hermitian operator on the Hilbert space $\mathcal{H}_E$ of  apparatus $E$. The  spectral decomposition of $H(n)$  is given by }
\begin{equation}\label{2020_12_17_1}
  H(n)=\sum_{\alpha}h_{\alpha}(n)|\sigma_\alpha(n)\rangle \langle \sigma_\alpha(n)|~,
\end{equation}
{where $|\sigma_\alpha(n)\rangle$ is eigenstate of $H(n)$ with eigenvalue $h_{\alpha}(n)$.}

Because of $X=[A,V]$,
one can find $x_{nn}=0$. Thus, solving the integral equation\eqref{2020_10_22_4} with the
initial condition\eqref{2020_10_22_5}, we obtain
\begin{equation}\label{2020_10_22_6}
  X(k\tau_m)=\sum_{n\neq l}x_{nl}e^{iH(n) k\tau_m}e^{-iH(l) k\tau_m}|s_n\rangle\langle s_l|~.
\end{equation}
Finally, substituting Eq.\eqref{2020_10_22_6} into Eq.\eqref{2020_10_20_8}, we obtain $A(t_N)$
as (for more details see Appendix A)
\begin{equation}\label{2020_10_22_7}
  \begin{split}
   A(t_N) &=-i\tau\sum_{n\neq l,\alpha,\beta} \frac{\sin[(h_\alpha(l)-h_\beta(n))N\tau_m/2]}{\sin[(h_\alpha(l)-h_\beta(n))\tau_m/2]}\times \\
      & ~~~~ x_{nl}y_{\beta\alpha}(nl){e^{-i(h_\alpha(l)-h_\beta(n))(N-1)\tau_m/2}\times}\\
     &~~~~ |s_n,\sigma_\beta(n)\rangle \langle s_l, \sigma_\alpha(l)|+A(t_0)+o(\tau)~,
  \end{split}
\end{equation}
where $y_{\beta\alpha}(nl)=\langle \sigma_\beta(n)|\sigma_\alpha(l)\rangle$.

From the above expression of $A(t_N)$, one can see that when
\begin{equation}\label{2020_12_1_1}
  { \sin\bigg[\big(h_{\alpha}(l)-h_{\beta}(n)\big)\frac{\tau_m}{2}\bigg]\neq 0~,}
\end{equation}
$A(t_N)$ will
tend to $A(t_0)$ as $N$ tends to infinity. In order to satisfy this condition, we additionally require: \romannumeral1) {$H_M$} is not degenerate, namely, $h_{\alpha}(l)\neq h_{\beta}(n)$ and \romannumeral2)
\begin{equation}\label{2020_12_1_2}
  \tau_m\neq
\frac{2k\pi}{h_{\alpha}(l)-h_{\beta}(n)}~,
\end{equation}
where $k$ is an integer and {$n\neq l$}.
Thus, when these two requirements are both satisfied, the QZE phenomenon will occur  according to the arguments in the last section. Set
\begin{equation}\label{2020_12_1_3}
  \tau_m^*=
\frac{2k\pi}{h_{\alpha}(l)-h_{\beta}(n)}~.
\end{equation}
Hence, when $\tau_m=
\tau_m^*$ but  $\Lambda_S$ is not degenerate, $A(t_N)$
can not still tend to $A(t_0)$ as $\tau$ tends to zero. Thus, $\tau_m^*$
is a critical measurement time, which QZE phenomenon can not occur when the value of $\tau_m$ equals $\tau_m^*$. Since our approach does not depend on a concrete physical model, it is concluded that the critical measurement time $\tau_m^*$ quite universally exists in the QZE phenomenon. One can note that the magnitude of $\tau_m^*$ only depends on the energy spectral structure of the couplings of the  system $S$ to the apparatus $E$, due to the assumption of QND measurement.

The effect of critical measurement time was firstly demonstrated by Xu \textit{et al.}, where they calculated the average photon number after $N$ successive measurements for a cavity-QED model and found the photon number could not be suppressed when $\tau_m$ took some special values \cite{Xu_PRA_2011}. Later,  the critical measurement time of a spin model was calculated by Zheng \textit{et al.}, and they verified it via a nuclear magnetic resonance experiment \cite{Zheng_PRA_2013_QZE}. Thus, the critical measurement time was calculated model by model previously. Now, we obtain a model-free expression of the critical measurement time, and prove it can universally exist in the practical QZE.
Moreover, since the WPC does not involve the duration $\tau_m$ of measurement, the effect of critical measurement time can not be explained in terms of the WPC. Thus, the effect of critical measurement time is a significant prediction of the dynamical approach, which largely differs from the prediction of the WPC.

It is pointed out that the expression of critical measurement time $\tau_m^*$ must be also obtained in the Schr\"{o}dinger picture, due to the  equivalence of the two pictures. Thus, we give a complete derivation for $\tau_m^*$ in the  Schr\"{o}dinger picture in  Appendix B.


\section{\label{sec_2020_10_22_4} XX model}

Consider a chain of $L$ two-level systems with the Hamiltonian
\begin{equation}\label{2020_10_23_1}
  H=\sum_{n=0}^{L-1}\frac{1}{2}\omega \sigma_n^z+g(\sigma_n^+\sigma_{n+1}^-+\sigma_n^-\sigma_{n+1}^+)~.
\end{equation}
Here, $\sigma_n^+=(\sigma_n^-)^\dagger=|e\rangle_n\langle \mathrm{g}|$, $\sigma_n^z=|e\rangle_n\langle e|-
|\mathrm{g}\rangle_n\langle \mathrm{g}|$; $|e\rangle_n$ and $|\mathrm{g}\rangle_n$ are excited and ground states respectively
in the $n$-th site of this chain. This is the XX model \cite{Sachdev_QPT_2011}, and could be exactly solvable.

We wish to measure the excited population of site-0. Then the  Hamiltonian for
measurement reads as
\begin{equation}\label{2020_10_23_2}
  H_M=\sigma_0^+\sigma_0^-\otimes \Sigma_E~,
\end{equation}
where $\Sigma_E$ is an operator on the Hilbert space of apparatus $E$.

To consider time evolution of the chain, we use the Jordan-Wigner transform,
\begin{equation}\label{2020_10_23_3}
  \sigma_n^z=2c_n^\dagger c_n-1~,~~~\sigma_n^+=c_n^\dagger\prod_{i=0}^{n-1}(-\sigma_i^z)~,
\end{equation}
to get a free spinless fermion  Hamiltonian,
\begin{equation}\label{2020_10_23_4}
  H=\sum_{n=0}^{L-1}\omega c^\dagger_n c_n+g(c_n^\dagger c_{n+1}+c_{n+1}^\dagger c_n)~.
\end{equation}
Meanwhile, the initial Hamiltonian for measurement is transformed into
\begin{equation}\label{2020_10_23_5}
  H_M=c_0^\dagger c_0\otimes \Sigma_E~.
\end{equation}

Applying the Fourier transform
$c_n=\frac{1}{\sqrt{L}}\sum_{k=0}^{L-1}$$e^{i\frac{2\pi}{L}nk}b_k$,
we  write down the simple diagonal form from Eq.\eqref{2020_10_23_4}
\begin{equation}\label{2020_10_23_6}
  H=\sum_{k=0}^{L-1}\varepsilon_k b^\dagger_k b_k~,
\end{equation}
where $\varepsilon_k=\omega+2g\cos(2\pi k/L)$. Thus, we obtain the time evolution of the element of single particle reduced density matrix
\begin{equation}\label{2020_10_23_7}
  \begin{split}
    \langle c^\dagger_n c_m\rangle(t) & =\sum_{k,q=0}^{L-1}\frac{1}{L}e^{i\frac{2\pi}{L}mq-i\frac{2\pi}{L}nk} e^{i(\varepsilon_k-\varepsilon_q)t}\langle b_k^\dagger b_q\rangle(0) \\
      & =\sum_{x,y=0}^{L-1}f(n,m,x,y,t) \langle c^\dagger_x c_y\rangle(0)~,
  \end{split}
\end{equation}
which is also the lattice correlation function. Here,
\begin{equation}\label{2020_10_23_8}
\begin{split}
  f(n,m,x,y,t) & =\sum_{k,q=0}^{L-1}\frac{1}{L^2}e^{i\frac{2\pi}{L}(m-y)q-i\frac{2\pi}{L}(n-x)k}\times \\
    & ~~~~e^{i(\varepsilon_k-\varepsilon_q)t}~.
\end{split}
\end{equation}
Similarly, the dynamics of measurement evolution  is also written as
\begin{equation}\label{2020_10_23_9}
    \langle c_n^\dagger c_m\rangle(t)= \langle c_n^\dagger c_m\rangle(0)g(n,m,t)~,
\end{equation}
where
\begin{equation}\label{2020_10_23_10}
  g(n,m,t)=\left\{\begin{split}
                    & e^{i\Sigma_Et},~~~n=0,m\neq0\\
                    & e^{-i\Sigma_Et},~~~m=0,n\neq0 \\
                     &1,~~~\mathrm{others}
                \end{split}
  \right.~.
\end{equation}
 \begin{figure}[htbp]
\centering
\subfigure{
\centering
\includegraphics[width=3in]{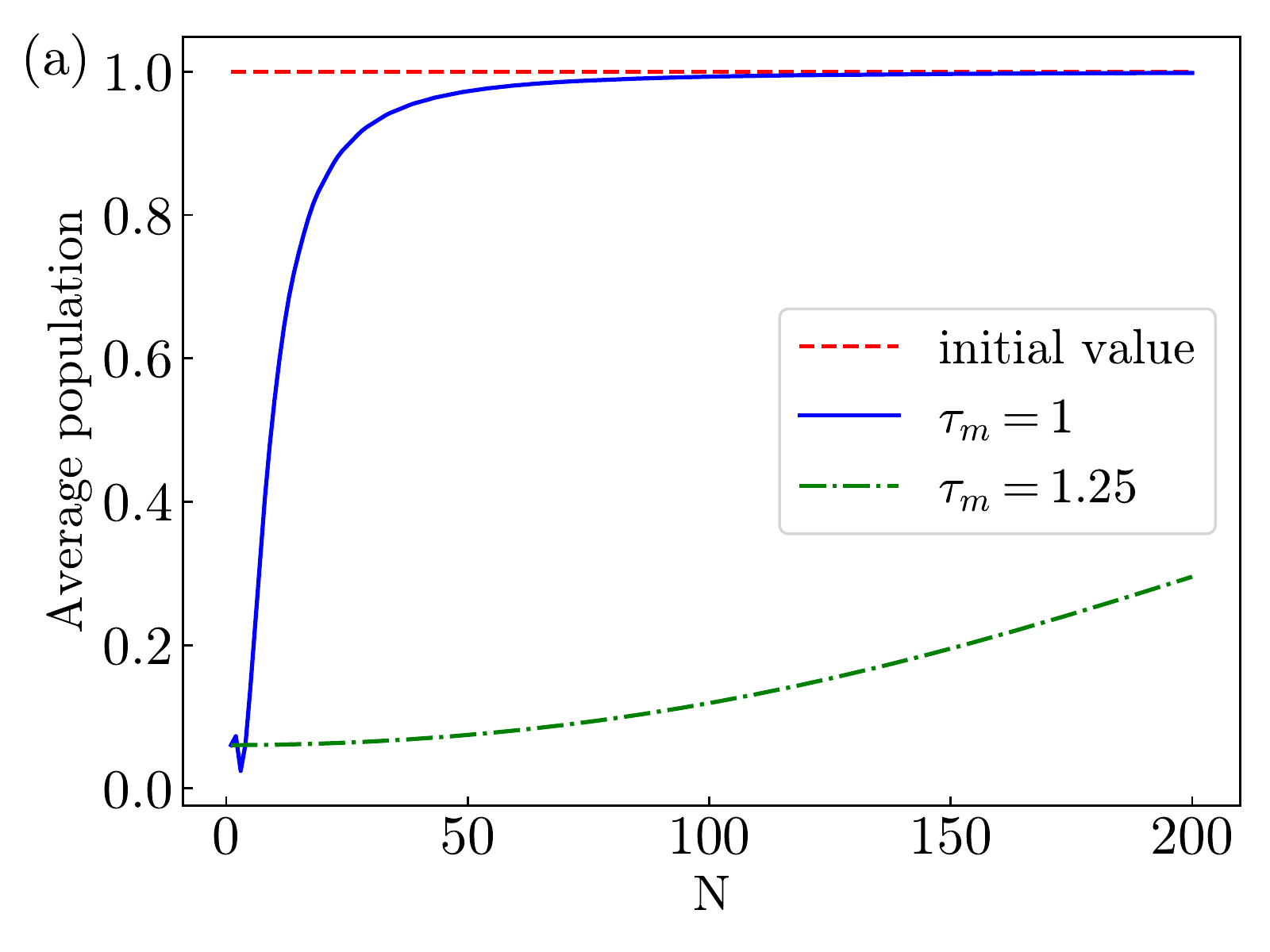}\label{Fig:expectation_2020_10_24_a}
}%

\subfigure{
\centering
\includegraphics[width=3in]{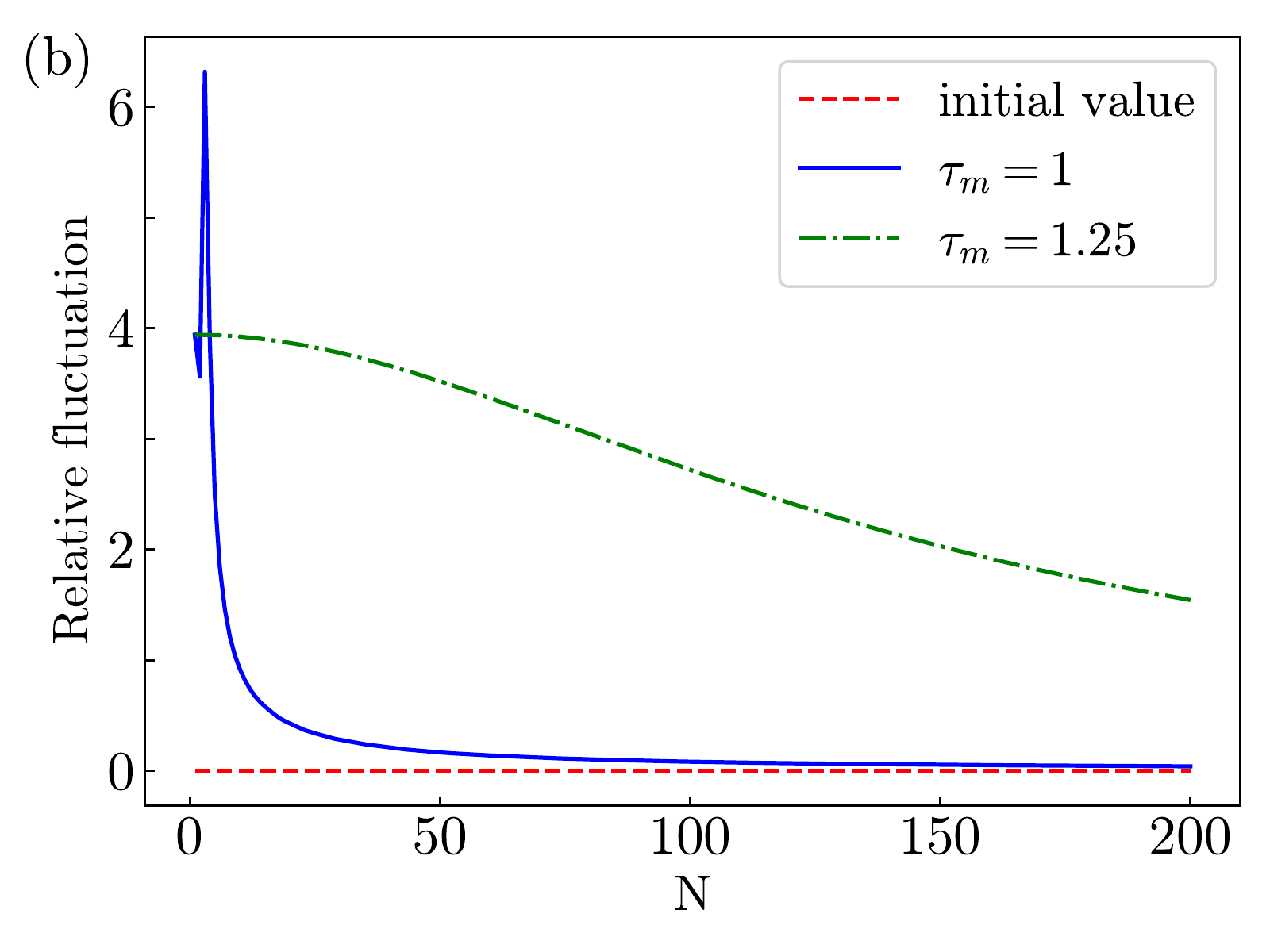}\label{Fig:expectation_2020_10_24_b}
}%

\centering
\caption{ \label{Fig:expectation_2020_10_24}  Average population (a) and its relative fluctuation (b) as a function of measurement number $N$. We fix $t=N\tau=1$, and choose
$r_E=g=5$, $L=30$. The average population and relative fluctuation rapidly tend to a constant and zero (red dashed line), respectively, when $\tau_m=1$ (blue solid line). But, this doesn't occur when $\tau_m$ is chosen specifically at $1.25$ (green dash-dotted line), which is close to a
critical measurement time.}
\end{figure}
According to Eq.\eqref{2020_10_23_7} and Eq.\eqref{2020_10_23_9}, the recursion relation of $ \langle c_n^\dagger c_m\rangle(t_k)$ and
$ \langle c_n^\dagger c_m\rangle(t_{k-1})$ is expressed as
\begin{equation}\label{2020_10_23_11}
   \begin{split}
     \langle {c}_n^\dagger{c}_m\rangle(t_k) & =\langle {c}_n^\dagger{c}_m\rangle(t_{k-1}+\tau)g(n,m,\tau_m) \\
      &=\sum_{x,y=0}^{L-1} F(n,m,x,y,\tau,\tau_m)\langle {c}_x^\dagger{c}_y\rangle(t_{k-1})~,
  \end{split}
\end{equation}
where $F(n,m,x,y,\tau,\tau_m)=f(n,m,x,y,\tau)g(n,m,\tau_m)$.  We arrange $ \langle {c}_n^\dagger{c}_m\rangle(t_k)$ into a column vector in order of $00,01,\cdots,0L-1,10,\cdots,L-1L-1$, and denote this vector by $C(t_k)$. Similarly,
we arrange $F(n,m,x,y,\tau,\tau_m)$ into a matrix in the same order, and denote it by $F(\tau,\tau_m)$. Thus, we  rewrite Eq.\eqref{2020_10_23_11} as
\begin{equation}\label{2020_10_23_12}
  C(t_k)=F(\tau,\tau_m)C(t_{k-1})~.
\end{equation}
According to Eq.\eqref{2020_10_23_12}, we obtain the relation between $C(t_N)$ and
$C(t_0)$,
\begin{equation}\label{2020_10_23_13}
  C(t_N)=F(\tau,\tau_m)^NC(t_{0})~.
\end{equation}

Assuming that the initial condition of the chain is $ \langle {c}_0^\dagger{c}_0\rangle(t_0)=1$, and $\langle {c}_m^\dagger{c}_n\rangle(t_0)=0$ for other $m,n$, which means that the initial state of site-0 is in excited state, and  remaining sites are in ground state. The initial state of apparatus part is the eigenstate of $\Sigma_E$ with eigenvalue $r_E$. Under this initial condition, we  write
\begin{equation}\label{2020_10_23_14}
  \langle {c}_0^\dagger{c}_0\rangle(t_N)=F(\tau,\tau_m)^N(00,00)~,
\end{equation}
where $F(\tau,\tau_m)^N(00,00)$ is denoted by the first row and the first column element of matrix $F(\tau,\tau_m)^N$. Because of $(c_0^\dagger c_0)^2=c_0^\dagger c_0$, the relative fluctuation is expressed as
\begin{equation}\label{2020_10_23_15}
 \frac{\Delta \langle {c}_0^\dagger{c}_0\rangle(t_N) }{\langle {c}_0^\dagger{c}_0\rangle(t_N)}=\sqrt{\frac{1}{\langle {c}_0^\dagger{c}_0\rangle(t_N)}-1}~.
\end{equation}

In this case, it is difficulty to give a specific analytic expression of $F(\tau,\tau_m)^N(00,00)$, so the critical measurement time can not be directly obtained  from $F(\tau,\tau_m)^N(00,00)$. However, according to the expression of critical measurement time,  we directly write
\begin{equation}\label{2020_10_23_16}
  \tau_m^*=\frac{2k\pi}{r_E}~,
\end{equation}
where $k$ is an integer. Thus, when $\tau_m\neq2k\pi/r_E$ and $\tau$ tends to zero, $ \langle {c}_0^\dagger{c}_0\rangle(t_N)$ will tend to $ \langle {c}_0^\dagger{c}_0\rangle(t_0)$, and $\Delta \langle {c}_0^\dagger{c}_0\rangle(t_N) /\langle {c}_0^\dagger{c}_0\rangle(t_N)$ will tend to zero.

The numerical results of average population and its relative fluctuation on site-0 are shown in Fig.\ref{Fig:expectation_2020_10_24_a} and Fig.\ref{Fig:expectation_2020_10_24_b}. In our calculation,
 $t=N\tau=1$, $r_E=g=5$, $L=30$ are used. For measurement time $\tau_m$, we choose two values $1$ (blue line) and $1.25$ (green line), which the later is close to the critical time $2\pi/r_E$. When $\tau_m=1$, average population and relative fluctuation rapidly tend one and zero, respectively. Thus, we know QZE phenomenon occurs. However,
 when $\tau_m$ is chosen at $1.25$, average population does not rapidly tend to one, so does not relative fluctuation. These numerical results are in agreement with our prediction about the
 universal effect of critical measurement time.

\section{\label{sec_2020_10_22_5}Conclusion}

In conclusion, the QZE phenomenon has been revisited in Heisenberg picture. Here, when the expectation value of an observable
and its fluctuation turn into a constant and zero respectively in the large-$N$ limit, we judge that the QZE phenomenon happens. With this observation,  we present a universal proof
for QZE in Heisenberg picture. Moreover, by studying the QZE in Heisenberg picture, we predict that it is universal that there exists critical measurement time $\tau_m^*$ and give a general expression for it. The expression of $\tau_m^*$ does not depend on a concrete physical model, but the magnitude of $\tau_m^*$ depends on the energy spectral structure of the couplings of the  system $S$ to the apparatus $E$, due to the assumption of QND measurement. This is a generalization of previous results, since the critical measurement time is only discussed in some concrete examples before. We also derive the expression of $\tau_m^*$ in Schr\"{o}dinger picture, due to  the equivalence of the two pictures. It is pointed out that the prediction about critical measurement time can not be explained by the WPC, due to the disappearance of the interaction duration $\tau_m$ in the WPC. Thus, the existence of critical measurement time is the key criterion to distinguish experimentally which is correct between the WPC and the dynamic method.  Our numerical results in XX model are in agreement with the prediction about the universal critical measurement time.

\section*{Acknowledgements}

Wu Wang is grateful to Jinfu Chen for helpful discussions. This work is supported by the National Basic Research Program of China (Grants No. 2016YFA0301201), NSFC (Grants No. 12088101,  No. 11534002), and NSAF  (Grants No. U1930403, No. U1930402).

\appendix
\section{The expression of $A(t_N)$}

Because of Eq.\eqref{2020_10_20_8} and the discussion in the Sec. \ref{sec_2020_10_22_3}, $A(t_N)$ is given by
\begin{equation}\label{2021_1_3_1}
  A(t_N)=-i\tau\sum_{k=0}^{N-1}X(k\tau_m)+A(t_0)+o(\tau)~.
\end{equation}
According to Eq.\eqref{2020_12_17_1} and Eq.\eqref{2020_10_22_6}, $X(k\tau_m)$ is rewritten by
\begin{equation}\label{2021_1_3_2}
\begin{split}
   X(k\tau_m) & =\sum_{n\neq l,\alpha,\beta} x_{nl}y_{\beta\alpha}(nl) e^{-i(h_\alpha(l)-h_\beta(n))k\tau_m}\times \\
    & ~~~~|s_n,\sigma_\beta(n)\rangle \langle s_l, \sigma_\alpha(l)|~,
\end{split}
\end{equation}
where $y_{\beta\alpha}(nl)=\langle \sigma_\beta(n)|\sigma_\alpha(l)\rangle$.

Substituting Eq.\eqref{2021_1_3_2} into Eq.\eqref{2021_1_3_1}, we obtain
\begin{equation}\label{2021_1_3_3}
  \begin{split}
    A(t_N) & = -i\tau\sum_{k=0}^{N-1}\sum_{n\neq l,\alpha,\beta} x_{nl}y_{\beta\alpha}(nl) e^{-i(h_\alpha(l)-h_\beta(n))k\tau_m}\times\\
      & ~~~~|s_n,\sigma_\beta(n)\rangle \langle s_l, \sigma_\alpha(l)|+A(t_0)+o(\tau)\\
      &=-i\tau\sum_{n\neq l,\alpha,\beta} \frac{\sin[(h_\alpha(l)-h_\beta(n))N\tau_m/2]}{\sin[(h_\alpha(l)-h_\beta(n))\tau_m/2]}\times \\
      & ~~~~ x_{nl}y_{\beta\alpha}(nl){e^{-i(h_\alpha(l)-h_\beta(n))(N-1)\tau_m/2}\times}\\
     &~~~~ |s_n,\sigma_\beta(n)\rangle \langle s_l, \sigma_\alpha(l)|+A(t_0)+o(\tau)~.
  \end{split}
\end{equation}
Therefore, we obtain the expression of $A(t_N)$.
\section{Critical measurement time in Schr\"odinger picture}
We can also derive critical measurement time in the Schr\"odinger picture. \(U(\tau)\) is defined as the unitary evolution operator of system \(S\) with the "measurement" turned off, and \(M(\tau_m)\) is the unitary measurement operator, we have
\begin{equation}
    U(\tau) = \ee^{- \ii \tau H} , \quad M(\tau_m) = \ee^{- \ii \tau_m H_M}
\end{equation}
and \(U(\tau,\tau_m)\) is evolution operator of the whole process with \(N\) frequent repeated measurements
\begin{equation}
    U(\tau,\tau_m) = \qty[M(\tau_m) U(\tau) ]^N~.
\end{equation}

 We rewrite \(U(\tau,\tau_m)\) by \(U_k(\tau) \equiv M(\tau_m)^k U(\tau) M(\tau_m)^{-k} \) as an multiproduct
\begin{equation}
    U(\tau,\tau_m) = \qty[\prod_{k=1}^N U_k(\tau)] M(\tau_m)^N~.
    \label{eq:U_c_trans}
\end{equation}
For short \(\tau\) or large \(N\), we expand \(U_k(\tau)\) to the first order of \(\tau\)
\begin{equation}
\begin{split}
    U_k(\tau) &  \approx 1 - \ii \tau M(\tau_m)^k H M(\tau_m)^{-k} \equiv 1 - \ii \tau H_k~,
\end{split}   \label{eq:U_k_expand}
\end{equation}
where $H_k=M(\tau_m)^k H M(\tau_m)^{-k}$.
The explicit form of $H_k$'s matrix elements is given as
\begin{equation}
    H_k = H_d+ \sum_{m\neq n} H_{mn} \ee^{- \ii k \tau_m H(m)} \ee^{\ii k \tau_m H(n)} \dyad{s_m}{s_n}~,
    \label{eq:H_k_dig-off-diag}
\end{equation}
where $H_d$ is the diagonal term of $H$.
Substituting Eq.\eqref{eq:U_k_expand}, Eq.\eqref{eq:H_k_dig-off-diag} and Eq.\eqref{2020_10_22_5} into Eq.\eqref{eq:U_c_trans}, we obtain
\begin{equation}
    U(\tau,\tau_m) \approx \qty(1 - \ii t H_d - \ii \frac{t}{N} \Gamma) M(\tau_m)^N~,
\end{equation}
where
\begin{equation}
    \begin{split}
       \Gamma & = \sum_k \sum_{m\neq n} H_{mn} \ee^{-\ii k \tau_m H(m)} \ee^{\ii k \tau_m H(n)} \dyad{s_m}{s_n}\\
       & = \sum_{m\neq n } \Lambda_{mn} H_{mn} \dyad{s_m}{s_n}~,\\
       \Lambda_{mn}  & =  \sum_{k=1}^{N} \ee^{-\ii k \tau_m H(m)} \ee^{\ii k \tau_m H(n)}~.
    \end{split}
\end{equation}
Taking Eq.\eqref{2020_12_17_1} into account, we obtain
\begin{equation}
    \ee^{-\ii k \tau_m H(m)} = \sum_{\alpha} \ee^{- \ii k \tau_m h_{\alpha}(m) } \dyad{\sigma_{\alpha}(m)}~.
\end{equation}
Thus
\begin{equation}
    \begin{split}
            \Lambda_{mn} &= \sum_{k=1}^{N} \sum_{\alpha\beta} y_{\alpha\beta}(mn)\ee^{-\ii k \tau_m h_{\alpha}(m)} \ee^{\ii k \tau_m h_{\beta}(n)}\times\\
           & ~~~~ \dyad{\sigma_{\alpha}(m)}{\sigma_{\beta}(n)}\\
            &= \sum_{\alpha\beta} y_{\alpha\beta}(mn) \qty[\sum_{k=1}^{N}\ee^{-\ii k \tau_m (h_{\alpha}(m) - h_{\beta}(n) )}] \times\\
          &  ~~~~\dyad{\sigma_{\alpha}(m)}{\sigma_{\beta}(n)}~,
    \end{split}
\end{equation}
where \(y_{\alpha\beta} = \braket{\sigma_{\alpha}(m)}{\sigma_{\beta}(n)}\). If \(\ee^{-\ii \tau_m (h_{\alpha}(m) - h_{\beta}(n) )} \neq 1\) the summation over \(k\) is
\begin{equation}
\begin{split}
   \sum_{k=1}^{N}\ee^{-\ii k \tau_m (h_{\alpha}(m) - h_{\beta}(n) )} &   = \frac{\sin[\frac{1}{2} \tau_m N (h_\alpha (m) - h_\beta (n) ) ]}{\sin[\frac{1}{2}\tau_m (h_\alpha(m) - h_\beta (n))]}\times \\
    & ~~~~ \ee^{-\ii \tau_m (N+1) (h_\alpha(m) - h_\beta (n)) /2} ~.
\end{split}
\end{equation}
The above equation tells us that $\Lambda_{mn}$ is finite when \(\ee^{-\ii \tau_m (h_{\alpha}(m) - h_{\beta}(n) )} \neq 1\) . Thus, we have
\begin{equation}\label{eq:summationoverk}
    \lim_{N\rightarrow \infty} \frac{\Gamma}{N}  = 0~.
\end{equation}
This means the vanishing of off-diagonal terms of unitary evolution operator due to the frequent measurement, or there is QZE in other words. However, when
\begin{equation}
    \tau_m = \frac{2 l\pi}{h_{\alpha}(m) - h_{\beta}(n)} \equiv \tau_m^{*}
\end{equation}
which means \(\ee^{-\ii \tau_m (h_{\alpha}(m) - h_{\beta}(n) )} = 1\), Eq.\eqref{eq:summationoverk} not hold anymore, the off-diagonal terms of evolution operator remains when \(N\rightarrow \infty\), the QZE can not occur, making \(\tau_m^*\) the critical measurement time.

\bibliographystyle{unsrt}
\bibliography{bib/ref}

\end{document}